\documentclass[aps,prb,12pt,notitlepage,citeautoscript,superscriptaddress]{revtex4-1}

\usepackage[]{hyperref} 
\hypersetup{pdfauthor={Shalaev, Walasik, Litchinitser},pdftitle={Optically tunable topological photonic crystal}}
\usepackage{amsmath}
\usepackage{amsfonts}
\usepackage{textcomp}
\usepackage[]{graphicx}
\usepackage{epstopdf}
\usepackage{mathrsfs}
\usepackage{amsbsy}
\usepackage{cleveref}
\usepackage{lipsum}
\usepackage{nicefrac}
\usepackage{color}

\usepackage[utf8]{inputenc}
\usepackage[greek,english]{babel}
\usepackage{CJKutf8}

\usepackage{xr}

\usepackage{url}

\crefformat{equation}{Eq.~(#2#1#3)}
\crefformat{figure}{Fig.~#2#1#3}

\Crefformat{equation}{Equation~(#2#1#3)}
\Crefformat{figure}{Figure~#2#1#3}

\graphicspath{{./figures/}}

\begin{document}

\title{Optically tunable topological photonic crystal}

\author{Mikhail I. Shalaev}
\affiliation{Department of Electrical and Computer Engineering, Duke University, Durham, North Carolina 27708, USA}
\author{Wiktor Walasik}
\affiliation{Department of Electrical and Computer Engineering, Duke University, Durham, North Carolina 27708, USA}
\author{Natalia M. Litchinitser}
\affiliation{Department of Electrical and Computer Engineering, Duke University, Durham, North Carolina 27708, USA}

\date{\today}

\begin{abstract}
Topological photonic insulators pave the way toward efficient integrated photonic devices with minimized scattering losses. 
Optical properties of the majority of topological structures proposed to date are fixed by design such that no changes to their performance can be made once the device has been fabricated. 
However, tunability is important for many applications, including modulators, switches, or optical buffers. 
Therefore, we propose a straightforward way for the dynamic control of transmission in a silicon-based topological photonic crystal enabled by all-optical free-carrier excitation allowing for fast refractive-index modulation. 
The changes in both real and imaginary parts of refractive index cause up to 20~nm blue-shift of the transmission spectrum and approximately 85\% reduction in transmission.
With the control mechanism used here, switching times of the order of nanoseconds can be achieved.
The structures proposed here are compatible with the standard semiconductor industry fabrication process and operate at telecommunication wavelengths.  
\end{abstract}

\maketitle

Topological insulators (TIs), that were first discovered in condensed matter physics~\cite{kane05a,kane05,moore10,Bernevig13,fereira13,katmis16,Jotzu}, support conduction on their boundaries but behave as insulators in the interior.
Importantly, the energy transport on the edges is topologically protected and robust against structural perturbations and disorder.
Recent progress in engineered materials, such as metamaterials and artificially-created crystals has laid the foundation for the development of optical~\cite{umucaliar,hafezi11,Khanikaev12,fang12,hafezi13,Rechtsman13,Barik666} and acoustic~\cite{Lu16} analogs of TIs.
Topological photonic systems promise a new generation of chip-scale photonic devices and facilitating energy-efficient on-chip information routing and processing~\cite{lu14}. 
However, the demonstration of tunable topological photonic devices with on-demand control of light propagation remains a grand challenge.

Photonic TIs (PTIs) can be divided into two categories with broken or preserved the time-reversal (TR) symmetry.
The first experimental realization of anomalous-Hall PTI was performed in microwave frequencies using gyromagnetic material and magnetic field to break the TR symmetry~\cite{wang09}.
The gyromagnetic response of materials vanishes towards shorter wavelengths necessitating alternative approaches for the realization of PTIs in the visible and near-infrared ranges.
Another method to realize the PTIs with broken TR symmetry in a linear optical systems relies on temporal modulation of the structure with precise phase control~\cite{feng12}.
In structures with preserved TR symmetry, the inversion symmetry can be broken to realize less robust, but more feasible topological systems.
Recently, the Floquet TI was proposed where an array of evanescently-coupled waveguides was modulated in the propagation direction mimicking the temporal modulation of the structure~\cite{Rechtsman13}.
In another approach, artificial magnetic field was generated in an array of coupled ring resonators, where the phase accumulated by the electromagnetic wave during propagation around a unit cell was equivalent to that acquired by an electron moving in an external magnetic field~\cite{hafezi13}.
The other realizations of the PTIs include metamaterial structures~\cite{Khanikaev12} and photonic-crystal-based (PC-based) systems~\cite{wu15,Ma16,Chen17,Dong17,Barik666,shalaev19}.
As shown above, optics offers a unique platform for realizing many remarkable topological phenomena at room temperature and without strong magnetic field. 
Moreover, optical analogs of TIs bring topological phenomena to the domain of practical applications, including robust energy transport in compact, integrated photonic devices, all-optical circuitry, and communication systems. 
For many of these applications, dynamically-controlled scattering-free propagation of light is essential. 
However, nowadays, the majority of proposed PTIs operate in a fixed wavelength range and their mode of operation cannot be dynamically tuned. 
Recently, first steps toward the design and realization of tunable topological photonic structures have been made~\cite{desnavi18,li18,zhao18,saei18,Dobrykh18,Leykam18}.
Dynamic tunability was proposed in 1D and 2D structures using mechanism based on refractive-index change due to Kerr-type nonlinearity~\cite{Dobrykh18}, using thermal phase-changing material~\cite{li18} and by liquid crystal reorientation~\cite{desnavi18}.
Tunable TIs for elastic\cite{zhao18} and optical waves\cite{saei18} based on mechanical control of geometric parameters of stretchable structures were reported.

Here, we design and demonstrate tunable, on-chip, integrated PTI based on the valley-Hall effect in silicon PCs operating at the telecommunication wavelength~\cite{Collins16, Ma16}. 
Non-trivial topology of the crystal ensures backscattering-free light propagation around the path with four sharp turns and allows the structure to be immune against defects and imperfections. 
Tunability of the structure is enabled by the free-carrier (FC) excitation initiated by the pump beam and resulting in reduction of the real part of the refractive index and increase of the imaginary part of the refractive index. 
The refractive-index change in turn leads to the shift of the bandgap and, correspondingly, to the change in the transmission-peak position.

Previously, several conventional (trivial) tunable photonic devices based on the FC injection in semiconductors were demonstrated, including microcavities, integrated waveguides, and optical modulators~\cite{Handbook_Silicon_Photonics,Kampfrath2009,Opheij2013,Handbook_of_Optics, liu04,Bruck_2016}.
Free-carries can be injected electrically by applying electric current or optically by illuminating a semiconductor with light.
The electrical method requires complicated fabrication procedure due to the necessity of deposition of electrical contacts.
Therefore in this work, we focus on all-optical refractive-index manipulation by means of PC illumination with ultra-violet (UV) radiation.
This approach enables fast switching times of the order of nanoseconds corresponding to gigahertz (GHz) modulation frequency.
Furthermore, such Si-based structures are compatible with contemporary semiconductor industry fabrication processes.
Combining the concepts of topological protection and all-optical modulation may pave the way for future robust and dynamically controllable devices for optical communication.

\begin{figure}[!t]
	\includegraphics[width = \textwidth, clip = true, trim = {0 0 0 0}]{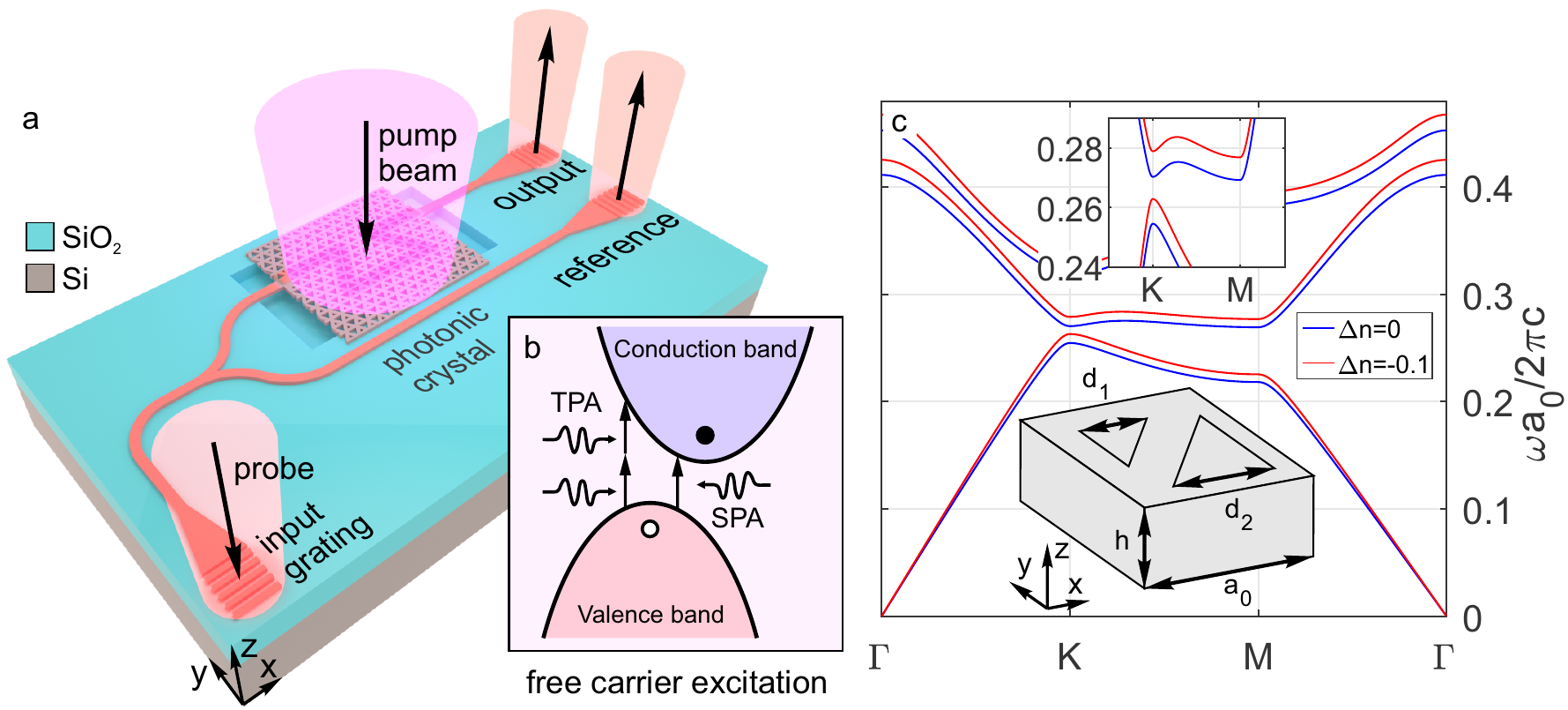}
	\caption{
		\textbf{Principle of operation.}
		Tuning of transmission of the topological photonic crystal is enabled by refractive index modulation (see panel \textbf{c}) due to optically-induced free-carrier excitation (see panel \textbf{b}).
		\textbf{a},~Near-infrared probe light was coupled to the chip by an input diffraction grating, then split in two parts where half goes to the topological photonic crystal and another part is used as reference. 
		The transmitted light is out-coupled from the chip by a pair of diffraction gratings. 
		Photonic crystal is illuminated by ultra-violet (UV) pump beam to control the refractive index of silicon.
		\textbf{b},~Two mechanisms of the free-carrier excitation:
		single-photon absorption (SPA) occurs when the photon energy exceeds the bandgap size and electrons are excited directly from the valence band to the conduction band;
		two-photon absorption (TPA), happens when two photons are absorbed simultaneously and the electron is excited to the conduction band.
		\textbf{c},~Band diagram for the hexagonal photonic crystal slab with two triangular holes per unit cell.
		The inset shows the geometry of the unit cell.
		The parameters used are: $a_0=423$~nm, $h=270$~nm, $d_1=0.4a_0$ and $d_2=0.6a_0$ and the effective refractive index of silicon is assumed to be $n_{\rm{Si,eff}} = 2.965$. 
		Illumination of the sample with a UV pump beam induces a refractive-index change in the semiconductor material.
		Assuming the index change of $\Delta n=-0.1$, the bandgap position is shifted towards higher frequencies (shorter wavelength).
		The inset shows the zoomed in picture of the band-diagram in the vicinity of the bandgap.
		The UV illumination can be used to control the spectral position of the bandgap and, correspondingly, the operation frequencies of the photonic crystal.}
	\label{fig:principle}
\end{figure}

We study tunability of topological PC fabricated on the standard silicon-on-insulator platform shown schematically in~\cref{fig:principle}a.
The sample characterization is enabled by diffraction gratings connected to the PC by Si-wire waveguides.
The light is coupled to the chip by the input grating and it is split in two parts. 
One part propagates through the topological PC and is out-coupled by the output grating. 
The second part is used as a reference. 
The refractive index of silicon is controlled by the UV pump beam illuminating the PC.

We consider a PC slab supporting the valley-Hall effect described in detail in Ref.~\cite{shalaev19}.
The unit cell of the PC contains two triangular holes in a silicon slab surrounded by air on the top and bottom. 
The symmetry in the $z$-direction allows us to consider solely transverse-electric-like (TE-like) modes in this work.
We used a standard approach to design topologically non-trivial structures. 
First, a system exhibiting a Dirac cone in its band structure was found. 
Second, certain symmetries of the original structure are broken in order to open a topological bandgap.
In our design, the Dirac cone exists when the triangular holes are of the same size.
For the dissimilar triangles, the inversion symmetry is broken; the symmetry of the structure is reduced from C$_6$ to C$_3$ and a non-trivial bandgap opens, as illustrated in~\cref{fig:principle}c.
The similarity between electronic and photonic crystals implies that many topological phenomena predicted for electronic systems should be observable in photonic structures with the same symmetry properties.
The structure studied here is similar to the boron-nitride structure that is well-known to support the valley-Hall effect for electrons.

We used UV light in order to enable efficient FC excitation~\cite{Opheij2013,Kampfrath2009} and induce refractive index change in Si.
Silicon has the bandgap size of around $E_g = 1.14$~eV, which is equal to photon energy at the wavelength of approximately 1.1~$\mu$m.
There are two processes that contribute to the FC excitation shown schematically in~\cref{fig:principle}b.
When the energy of the irradiation light exceeds the bandgap size of Si, electrons from the valence band can be directly excited to the conduction band, leading to a single-photon-absorption (SPA) process.
The second mechanism, that contributed to the FC excitation, is two-photon absorption (TPA), where electrons are excited to twice as high energy levels due to simultaneous absorption of two photons.
The SPA dominates at low pump fluence, while the TPA prevails at high power-levels~\cite{Sokolowski2000}.
It should be noted that other competing processes (in addition to the FC excitation) take place in semiconductors simultaneously, including the Kerr nonlinear index modulation, and FC dispersion effect~\cite{Handbook_Silicon_Photonics,Handbook_of_Optics}, whose contribution is negligible for the parameters used in our experiments.
We used the Drude model to describe the dielectric permittivity variation under light illumination~\cite{Sokolowski2000}:
\begin{equation}\label{eq:permittivity}
\Delta\epsilon_\textrm{FC}(F_\textrm{eff})=-\left[\frac{\omega_p(F_\textrm{eff})}{\omega}\right]^2\frac{1}{1+i\frac{1}{\omega\tau_D}}, 
\end{equation}
where $\omega$ is the  angular frequency of the incident light;
$\tau_D$ denotes the Drude damping time;
$F_\textrm{eff}$ is the effective pump fluence;
$\omega_p(F_\textrm{eff})=\sqrt{N_\textrm{e-h}(F_\textrm{eff})e^2/\epsilon_0~m_\textrm{opt}m_e}$ is the plasma frequency,
$e$ is the electron charge,
$m_\textrm{opt}=\left(m_e^{*-1}+m_h^{*-1}\right)^{-1}$ denotes the optical effective mass of carriers,
$m^*_{e,h}$ are the mobility effective masses of electrons and holes,
$m_e$ is the electron mass,
$N_\textrm{e-h}(F_\textrm{eff})$ is the electron--hole density depending on the pump fluence, and
$\epsilon_0$ stands for the vacuum permittivity.

\begin{figure}[!t]
	\includegraphics[width = \textwidth, clip = true, trim = {0 0 0 0}]{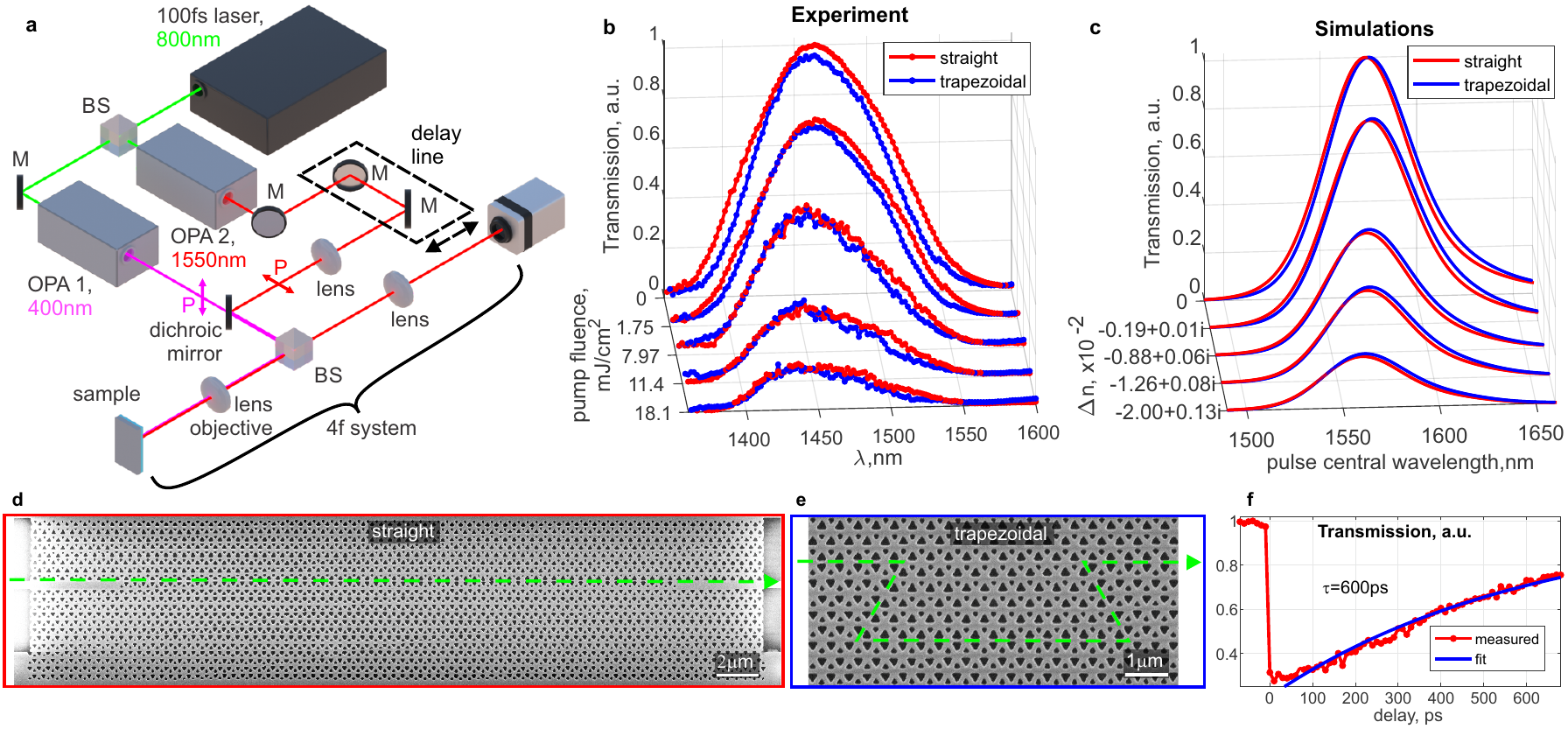}
	\caption{
		\textbf{All-optical transmission control for topological edge states.}
		\textbf{a},~Schematics of the experimental setup. 
		Light from a Ti:Sapphire laser is fed into two optical parametric amplifiers to generate the pump and probe beams.
		Relative arrival times of the pump (UV) and probe (NIR) pulses are controlled by a delay line.
		The two beams are combined by a dichroic mirror and routed towards the sample by a beam splitter. 
		Light is focused on the sample by an infinity-corrected objective lens; a 4f-system is used for sample imaging and transmission measurements.
		\textbf{b,c},~Experimentally measured and numerically simulated transmission spectra for various levels of the pump fluence.
		Refractive-index change in the numerical analysis was fitted assuming a Gaussian spatial distribution of the index with parameters corresponding to the pump-beam size used in experiments.
		The transmission-peak position shifts by 20~nm towards shorter wavelengths (blue shift) at the highest pump fluence while the peak transmission is reduced by around 85\%.
		The transmissions for the trapezoidal and straight paths are similar indicating lossless light propagation around sharp turns as a manifestation of the topological protection.
		\textbf{d,e},~Scanning electron microscopy images of two sample types measured in the experiment, having the straight edge and the one with four turns.
		Green dashed lines show the predicted light-propagation paths.
		\textbf{f},~Free-carrier-lifetime measurement based on the dependence of the normalized transmission on the delay between the pump and probe pulses.
		Red curve shows the measured data and the blue one shows the fitted transmission with free-carrier lifetime $\tau=600$~ps.
	}
	\label{fig:exp_vs_th}
\end{figure}

The electron-hole density is characterized by the equation~\cite{Sokolowski2000}:
\begin{equation}\label{eq:eh_density}
N_\textrm{e-h}(F_\textrm{eff})=\frac{2\pi F_\textrm{eff}}{h\omega}\left(\alpha+\beta\frac{F_\textrm{eff}}{2\sqrt{2\pi}t_0}\right),
\end{equation}
where $\alpha$ is the linear absorption coefficient,
$\beta$ denotes the TPA coefficient,
$t_0$ is the pump-pulse width, and $h$ denotes the Planck constant.
The model described by Eqs.~(\ref{eq:permittivity})~and~(\ref{eq:eh_density}) allows to estimate the Drude damping time and electron--hole density in Si, based on variation in the real and imaginary parts of the refractive index. 
The refractive index change at pump fluence of 18.1~mJ/cm$^2$ was found by comparison between the experimental and simulation results (see description of \cref{fig:exp_vs_th}) and allowed us to estimate the Drude damping time $\tau_D=10^{-14}$~s and electron-hole density $N_{\textrm{e-h}}=10^{19}~\textrm{cm}^{-3}$, which are typical values for Si~\cite{Bruck15,Sokolowski2000}.
The optical effective mass of carriers was assumed to be $m_{\textrm{opt}}=0.15$~\cite{Sokolowski2000}.
According to Eqs.~(\ref{eq:permittivity})~and~(\ref{eq:eh_density}) there are two parts contributing to the permittivity change: first is associated with the SPA and linearly depends on pump fluence $F_{\textrm{eff}}$; the second is responsible for the TPA with quadratic dependence on the UV radiation fluence $F_{\textrm{eff}}$.
The linear dependence of the permittivity change $\Delta \epsilon_{\textrm{FC}}$ on the pump fluence observed in our experiments is in agreement with the previous observations~\cite{Sokolowski2000,Bruck15,Opheij2013}, and allows to conclude that for low levels of the pump fluence used in our experiments, the SPA process prevails in the FC generation over the TPA. 

\begin{figure}[!t]
	\includegraphics[width = \textwidth, clip = true, trim = {0 0 0 0}]{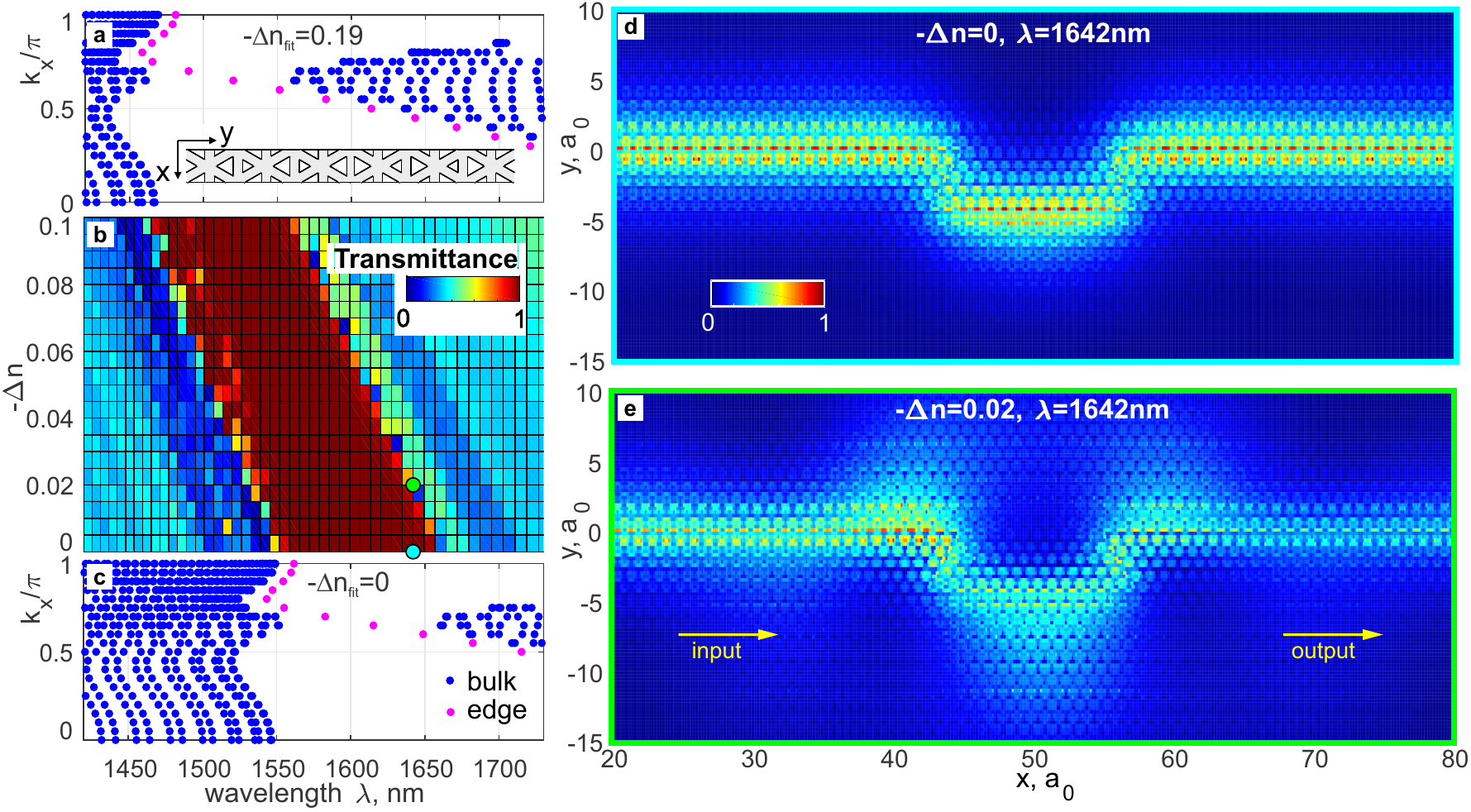}
	\caption{
		\textbf{Bandgap position control by photonic crystal illumination.}
		\textbf{b},~Transmittance spectrum for the trapezoidal-shaped interface in topological photonic crystal as a function of the refractive index change in silicon.
		For efficient guiding two conditions have to be satisfied: (i) the edge state must exist while (ii) no bulk states should be present at the guided frequency.
		\textbf{a,c},~Band diagrams for the super-cell periodic in the $x$-direction (see the inset) that contains two parts with different orientations of large and small triangles.
		The interface between the two parts is located in the middle of the structure (in $y$-direction).
		The position of the high-transmittance region is shifted towards shorter wavelengths upon reduction of the refractive index. 
		The width of the bandgap is also decreased due to the reduction of the index contrast.
		\textbf{d,e},~Energy-density distributions for pump illumination turned ON and OFF marked in panel \textbf{b} by the green and cyan dots, respectively.
		For the ON state, the transmittance is reduced at the wavelength $\lambda=1642$~nm compared to the off state.
	}
	\label{fig:T}
	
\end{figure}

Next, we study how the refractive index change $\Delta n$ induced by the FC excitation influences the light propagation in the PTI under investigation.
According to Eqs.~(\ref{eq:permittivity})~and~(\ref{eq:eh_density}) the refractive index change $\Delta n$ linearly depends on the fluence of the beam~$F_{\textrm{eff}}$, as for small refractive index perturbation $\Delta n \approx \Delta \epsilon/(2 n_{\rm{Si,eff}})$.
Therefore, the index change profile in the sample plane is assumed to have the same Gaussian distribution as the UV pump beam used in the experiment.
The transmittance, defined as a ratio of energy density after and before the four turns, as a function of the refractive index change and the wavelength is shown in~\cref{fig:T}b.
The region with high and nearly unitary transmittance corresponds to topologically protected light propagation with suppressed back-scattering.
A typical energy density distribution for the edge state is shown in~\cref{fig:T}d, where the light propagates around four turns without significant scattering resulting in a nearly perfect transmittance.
When the refractive index is decreased due to the FC excitation, the high-transmittance region shifts towards shorter wavelengths.
This behavior can be qualitatively explained by considering the changes in the band structure of the super-cell upon reduction of the refractive index, see Figs.~\ref{fig:T}a,~\ref{fig:T}c.
For this simulations the structure is periodic along $x$-direction and finite along $y$-axis with 20 unit cells in each region below and above the edge.
We show only the band diagram for positive values of the wave vector $k_x$. 
The band diagram is symmetric with respect to $k_x=0$ and another edge state can be found for $k_x<0$ that propagates in the opposite direction.
The scattering-free guiding occurs in the spectral region where a single topological edge state exists for each of the $K$- and $K'$-valleys. 
This guiding region in shown in green in the super-cell band diagrams shown in Figs.~\ref{fig:T}a and \ref{fig:T}c.
The refractive index is assumed to be uniform within the super-cell and its value was chosen such that the spectral position of the guiding-region matches the wavelength range of the high-transmittance obtained in the simulations shown in \cref{fig:T}b.

Figures~\ref{fig:T}d, \ref{fig:T}e show how the energy density distribution changes at single wavelength when the pump is turned on and the refractive index is reduced by $\Delta n = -0.02$.
Under the UV-light illumination, the bandgap is blue-shifted and there are bulk states supported at the studied wavelength $\lambda=1642~\textrm{nm}$.
As a result, the light scatters from the edge state into the interior of the crystal and the transmittance is reduced.
Finally, it should be mentioned that as the refractive index of silicon decreases, the index contrast between silicon and air is also reduced, resulting in a narrower bandgap.
This effect becomes notable for larger refractive index changes.

In order to confirm the theoretical predictions, samples with straight and trapezoidal-shaped interfaces were fabricated. 
The sample fabrication procedure is described in detail in the supplementary materials of our previous paper~\cite{shalaev19}.
The samples were measured with the experimental setup schematically shown in \cref{fig:exp_vs_th}a.
A Ti:Sapphire laser with the repetition rate of 1~kHz and 100~fs output-pulse width was used as a light source in the experiment.
Pulsed laser radiation was separated in two parts by a beam-splitter (BS) and routed into two optical parametric amplifiers (OPA), where the UV pump beam (400~nm), and the near-infrared (NIR) probe pulse (tunable near 1500~nm) were generated such that the pump and probe beams have orthogonal polarizations.
We used a delay line to control relative arrival times of the NIR and UV pulses.
The convex lens was employed to control the spot size of the pump beam in the sample plane.
The approximate size of the beam waist was $\sigma=37~\textrm{nm}$.
The pump and probe beams were combined together and directed onto the same propagation path by a dichroic mirror, and were focused on the sample by an infinitely corrected objective lens.
Another BS was used for routing light towards the sample.
The objective and an achromatic lens form a 4f-system that was used for sample imaging and transmission measurements.

\Cref{fig:exp_vs_th}b shows the dependence of the transmission spectra on the fluence of the pump beam measured for two samples: with a trapezoidal-shaped propagation path (blue lines) and with a straight interface with no bends (red lines). 
Upon the UV illumination, the refractive index decreases and the transmission peak shifts towards the shorter wavelengths (blue shift). 
Wavelength shifts of up to 20~nm have been measured for the highest pump-beam fluence.
Besides the reduction of the real part of refractive index, UV pump illumination results in a significant rise of absorption in silicon~\cite{Handbook_of_Optics,Handbook_of_Optics}. 
At the highest pump power, the peak transmission is reduced by approximately 85\%.

In the numerical simulations, we used the crystal with the same parameters as in the experiments.
We assumed that the real and imaginary parts of refractive index are linearly dependent on the pump-beam fluence and the pump beam was assumed to have a Gaussian profile.
First, the real part of refractive index was fitted to match the blue shift found in the experimental measurements. 
At the highest pump intensity, we obtain the refractive index change of $\textrm{Re}(\Delta n)=-0.02$.
Second, the absorption coefficient was calculated taking into account the group velocity in the medium found in edge-state simulations (see \cref{fig:T}c) and the measured transmission of around 15\% of the value with the pump beam switched off.
Then, the imaginary part of the refractive index was found to be $\textrm{Im}(\Delta n)=0.0013$ at highest pump beam power.
Thus, at the pump-beam fluence of 18.1 mJ/cm$^2$ the estimated refractive index change was $\Delta n=-0.02+0.0013i$.
Third, assuming the linear dependence of refractive-index change on the pump-beam power, the values of $\Delta n$ were calculated for other fluences used in the experiments.
The value of the index change was found to be comparable with the ones reported in the literature although it should be noted that it depends on several factors, such as, semiconductor doping, surrounding material, and patterning of the structure~\cite{Handbook_Silicon_Photonics,Handbook_of_Optics}.
The results of the numerical simulations are shown in \cref{fig:exp_vs_th}c, and they are in a good agreement with the experiment data.
As predicted, the transmissions for straight path and for the one with bends are similar, confirming that the energy transport stays robust even under strong UV illumination.

In order to estimate the minimum switching times attainable for the proposed structure, we characterized the times of the refractive-index recovery by measuring the FC lifetime.
The dependence of transmission on the delay time between the pump and probe pulses was measured.
The change of transmission due to the FC excitation can be characterized by the expression $T=1-A$, where $A$ denotes absorption, and assuming exponential dependence of absorption $A=A_0e^{-t/\tau}$ on pump-probe delay $t$, where $A_0$ is attenuation factor and $\tau$ is the FC lifetime.
By fitting the results shown in \cref{fig:exp_vs_th}f, we found the attenuation factor $A_0=0.794$ and the FC lifetime $\tau=600~\textrm{ps}$.
The lifetime observed here is comparable to those reported previously~\cite{Bruck_2016, Opheij2013}.
It is worth noting that the FC lifetime decreases as the pump-beam width $\sigma$ decreases~\cite{Opheij2013}. 
The FC lifetime measured here allows for the structure switching time in order of a nanosecond.
In the case when a faster modulation rate is required, the semiconductors with a direct bandgap~\cite{shcherbakov17} instead of silicon can be used. 
Alternatively, a control mechanism based on Kerr nonlinearity, allowing faster switching times~\cite{eggleton11,xu18}, can be employed.

In conclusion, we studied the all-optical modulation of silicon topological photonic crystal slab supporting the valley-Hall effect.
The non-trivial topology of the crystal ensures backscattering-free light propagation around the path with four sharp turns and allows the structure to be immune against defects and imperfections.
Switching of the crystal is enabled by the FC excitation by the pump beam causing the reduction of the real part of the refractive index and increase of absorption.
Consequently, the transmission peaks are blue-shifted by up to 20~nm and the transmission is reduced by around 85\%.
The control mechanism used here allows for switching at a gigahertz frequency.
Different methods of refractive index modulation can be explored, including phase-changing materials, electro-optical modulation and Kerr-nonlinearity.
For instance, if the third-order nonlinearity is employed, only the real part of the index can be modulated at even higher (femtosecond) rate, while the material absorption remains negligible.
The chalcogenide glasses might be a suitable platform for tunable topological photonic due to their transparency in the near infrared wavelength range, very low two-photon absorption, exceptionally high Kerr nonlinearity coefficient, and high linear refractive index allowing for strong light confinement on the nanoscale.
The system studied here is fully compatible with contemporary semiconductor fabrication techniques and operates at technologically important telecommunication wavelength.
This research paves the way for efficient and tunable photonic devices for future classical and quantum communication systems.

\section*{Methods}

We used the following definitions for transmittance and transmission of light.
Transmittance is the amount of light propagated through four bends in photonic crystal for the case of a continuous source.
Transmission is measured in the experiments and is computed in the numerical simulations assuming pulsed light source with the spectral width of 40~nm. 
Therefore, the transmission is the convolution of the transmittance and pulse spectra.
The simulations were performed in 2D approximation with effective refractive index $n_{\textrm{Si,eff}}=2.965$ (for switched off pump beam).
The plane wave expansion method was used to calculate the band diagrams shown in~\cref{fig:principle}.
The COMSOL Multiphysics Software was used for super-cell simulations and for transmission and transmittance calculations.
The peaks for simulated resulted are red-shifted with respect to the experimental results due to the discrepancy coming from the effective index approximation. 
The refractive index modification is assumed to have a Gaussian profile.
The value of $\Delta n$ in Figs.~\ref{fig:T}b, \ref{fig:T}d, \ref{fig:T}e and \ref{fig:exp_vs_th}c denotes half of the maximum index change, where the index change used in the simulation has a Gaussian distribution given by $2\Delta n \exp[-(r/\sigma)^2]$

\section*{Acknowledgments}
This work was supported by by Army Research Office (ARO) (W911NF-18-1-0348).

\section*{Data availability}
The data that support the plots within this paper and other findings of this study are available from the corresponding author upon reasonable request.

\bibliographystyle{naturemag}
\bibliography{vH_tunable_v6}

\section*{Author Contributions} 
M.I.S, W.W. and N.M.L proposed the initial idea. 
M.I.S., and W.W. designed and performed the analytical and numerical analysis of the structure. 
M.I.S. fabricated the sample, performed the experimental measurements, and analysed the results. 
M.I.S., W.W.,  and N.M.L. co-wrote the manuscript. 
N.M.L. supervised the work.

\section*{Competing interests} 

The authors declare no competing interests. 

\section*{Author Information} 

\textbf{Supplementary Information} is linked to the online version of the paper at \url{www.nature.com/nature}.

\textbf{Reprints and permissions information} is available at \url{www.nature.com/reprints}.

\textbf{Correspondence and requests for materials} should be addressed to N.M.L.

\end{document}